\documentstyle[prl,aps,twocolumn,epsf]{revtex}
\def\break#1{\pagebreak \vspace*{#1}}
\begin{document}

\draft

\title{Simple supersymmetric methods in neutron diffusion}

\author{H.-C. Rosu and J. Socorro
}

\address{
Instituto de F\'{\i}sica de la Universidad de Guanajuato, Apdo Postal
E-143, Le\'on, Guanajuato, M\'exico
}

\maketitle
\widetext

\begin{abstract}

We present the supersymmetric Witten and double Darboux (strictly isospectral)
constructions as applied to the
diffusion of thermal neutrons from an infinitely long line source.
While the Witten construction is just a mathematical scheme, the double Darboux
method introduces a one-parameter family of diffusion solutions which are
strictly isospectral to the stationary solution. They correspond to a
Darboux-transformed diffusion length which is flux-dependent.

\end{abstract}

\pacs{PACS numbers: 11.30.Pb, 28.20.Gd \hspace{1.5cm}
LAA number: physics/9612003 \hfill Nuovo Cimento B 114, 115-119 (1999)
}

\narrowtext


Supersymmetric one-dimensional (1D) quantum mechanics has been considered by
Witten in 1981
as a toy model for symmetry-breaking phenomena in quantum field theory
\cite{wi}. With great speed its status has changed to a powerful
research area as one can contemplate in the most recent
review \cite{rev}. In a series of papers, we obtained interesting results
for various
physical problems by using Witten's factorization procedure of the 
Schr\"odinger 1D operator and a more general
supersymmetric double Darboux method \cite{rs} introduced by Mielnik \cite{M}.
The aim of this work is to apply these two simple, supersymmetric
methods to the theory of diffusion of thermal neutrons.
We shall use the illustrative example of the neutron diffusion
problem as presented in the textbook of Arfken \cite{ar} referring to an
infinite line (Dirac delta) source of
neutrons, thus providing the Green's function for this case.
The steady state continuity equation for the neutrons reads
$$
D\nabla ^2\phi -\Sigma _{a} \phi +S=0~,
\eqno(1)
$$
where the first term represents the diffusion, the second stands for the
absorption losses and $S$ stands for the source strength. The diffusion
constant $D$ is related to the neutron mean free path $\lambda _{s}$
as follows
$D=\frac{\lambda _{s}}{3(1-2/3A)}$, where $A$ is the atomic number of the
scattering nucleus and enters as a correction for the anisotropy of
scattering in the laboratory system. The solution $\phi$ has the physical
meaning of neutron flux being the product of neutron density times
average velocity. Finally, $\Sigma _{a}$ is the macroscopic absorption
cross section, i.e., the product of the microscopic (atomic)
absorption cross section and the number of atoms per unit volume. Usually
it is assumed that the absorption is small compared to the scattering. As we
said, the neutron source is considered as an infinitely long line source that 
can be taken
along the $z$-axis embedded in an infinite diffusing medium. Its strength
is $S_{0}$ neutrons per unit length per unit time. Thus $S=S_{0}\delta (\rho)$
where $\delta (\rho)$ is the cylindrical Dirac delta function. For $\rho \neq 0$,
in view of the no $z$ and $\theta$ dependence,
one gets the radial diffusion equation
\break{1.3in}
$$
\rho ^2\frac{d^2\phi}{d\rho ^2}+\rho\frac{d\phi}{d\rho}-\rho ^2k^2\phi=0~,
\eqno(2)
$$
where we have made use of the ``diffusion length"
$k^{-1}=\sqrt{D/\Sigma _{a}}$. Then
one can write the general solution in terms of the modified Bessel functions
$\phi =a_1I_0(k\rho)+a_2K_0(k\rho)$. The physical solution is only the
$K_0$ term because the flux is supposed to decrease at large distances. 
The constant $a_2$ may be determined by requiring that $D$ times the integral
of the negative gradient of the neutron flux around the edge of a small
pillbox of unit height be equal to the production strength $S_0$ within the
pillbox.
The box is small ($\rho\rightarrow 0$) to eliminate absorption and gives
$$
S_0=\lim _{\rho\rightarrow 0}Da_2
\int [-\nabla K_{0}(k\rho)\cdot \rho _0] \rho d\theta ~,
\eqno(3)
$$
which is a two-dimensional form of Gauss's law. Using the series form of
$K_0(k\rho)$ this turns into
$$
S_0=Da_2\lim _{\rho \rightarrow 0}\frac{2\pi \rho}{\rho}~,
\eqno(4)
$$
or
$$
\phi =\frac{S_0}{2\pi D}K_0(k\rho)
\eqno(5)
$$
for the final form of the common solution.

Let us pass now to the supersymmetric constructions. We have already
presented these schemes in previous papers \cite{rs}.
One needs a self-adjoint form of Eq.~(2) that can be obtained by the change
of function $\phi=\rho ^{-1/2}\psi$ leading to
$$
\psi ^{''}-(k^2-\frac{1}{4\rho ^2})\psi=0~.
\eqno(6)
$$
If Eq. (6) is interpreted as
a Schr\"odinger equation at fixed zero energy with the potential
$V_{B}(\rho)=k^2-\frac{1}{4\rho ^2}$, then one can think of supersymmetric
quantum mechanical methods as follows.
 
 (i). Witten's construction has two steps. The first is the factorization of 
Eq. (6) by means
of the operators $A_1=\frac{d}{d\rho}+W(\rho)$ and
$A_2=\frac{d}{d\rho}-W(\rho)$, where $W$ is the so-called superpotential
function that can be determined from the initial (``bosonic")
Riccati equation $V_{B}=W^2-W^{'}$, or even more directly as the negative
of the logarithmic derivative of the stationary solution $\psi$
($W=-\frac{d}{d\rho}\ln \psi$). The second step
of the Witten construction means to pass
to a ``fermionic" problem by merely changing the sign of the derivative term
in the Riccati equation and determining the new potential $V_{F}$, which
is interpreted as a ``fermionic" partner of the initial potential. Thus,
$V_{F}=W^2+W^{'}$. This partner potential enters a ``fermionic" 
equation for which the factoring operators are applied in reversed order.
On the other hand, the superpotential should be the negative of
the logarithmic derivative of a ``nodeless" solution, which is
our case since modified Bessel functions of zero order occur. Moreover, the 
singularity at the origin of the superpotential function is not disturbing
because we have an infinite line source there. 

(ii). The double Darboux
construction in the supersymmetric framework allows one to use the general
solution $W_{gen}$ of the ``bosonic" Riccati equation and not the particular
one as for the Witten construction. Thus, the factorization operators are now
${\cal A}_{1}=\frac{d}{d\rho}+W_{gen}$ and
${\cal A}_{2}=\frac{d}{d\rho}-W_{gen}$.
In this way one can introduce a one-parameter
family of ``bosonic" potentials having the same ``fermionic" partner, i.e.
$$
V_{iso}=V_{B}-2\frac{d^2}{d\rho ^2}\ln({\cal I}(\rho)+\lambda)~,
\eqno(7)
$$
which is a {\em general} Darboux transform of $V_{B}$. Eq. (7) can be written
in the following form
$$
V_{iso}(\rho ,\psi ;\lambda)=
V_{B}(\rho)-\frac{4\psi \psi ^{'}}{{\cal I}(\rho)+\lambda}+
\frac{2\psi ^4}{({\cal I}(\rho)+\lambda)^{2}}~,
\eqno(8)
$$
where ${\cal I}(\rho)=\int _{0}^{\rho}\psi ^{2}(r)dr$ and $\lambda$ is
the family parameter, which is a real positive quantity, being as a matter
of fact the Riccati integration constant \cite{N}. Besides, there is
a modulational damping of the general solution, which reads
$$
\psi (\rho ;\lambda)= \sqrt{\lambda(\lambda+1)}
\frac{\psi (\rho)}{{\cal I}(\rho)+\lambda}~,
\eqno(9)
$$
where the square root is a normalization constant. One can check easily that
$W_{gen}=-\frac{d}{d\rho}\ln \psi (\rho; \lambda)$.
Some plots of the strictly isospectral potentials Eq.~(8) and the
isospectral solutions Eq.~(9) are presented in
Figures 1 and 2 for different $\lambda$ values and $a_1=a_2=1$;
the case $\lambda =\infty$
corresponds to the original ``bosonic" solutions.
The interesting point is that one can work with the general flux solution and
still get 
physically meaningful, localized-type solutions (see Fig.~3). We mention
that in quantum mechanics
irregular wavefunctions have been found useful in state
reconstruction \cite{L}.
To see the physical meaning of the isospectral construction for neutron
diffusion we make the change of function $\psi =\sqrt{\rho}\phi$ in the
Schr\"odinger equation $\psi ^{''}-V_{iso}\psi=0$ to get
$$
\rho ^2\frac{d^2\phi}{d\rho ^2}+\rho\frac{d\phi}{d\rho}-
\rho ^2k^2_{eff}\phi=0~,
\eqno(10)
$$
where
$k^{-1}_{eff}=[k^2-2\frac{d^2}{d\rho ^2}\ln({\cal I}(\rho)+\lambda)]^{-1/2}$
is an effective, flux-dependent diffusion length, which is the Darboux
transform of the constant diffusion length.
Thus, one can see the connection between the common radial neutron
diffusion equation and a similar isospectral diffusion.

In conclusion, we have shown that simple supersymmetric methods of 
1D quantum mechanics may provide interesting results in 
neutron diffusion. 

{\bf Acknowledgment}

This work was partially supported by the CONACyT projects 4868-E9406 and
3898P-E9607.


{\bf Figure captions}

{\bf Fig. 1}: 
 The original potential $V_{B}=k^2-\frac{1}{4\rho ^2}$ for
$k=1$ and four of the strictly isospectral potentials for 
$\lambda$ taking the values 1, 1000, 3000, 6000, respectively (the
corresponding wells are from left to right, respectively).

{\bf Fig. 2}: 
 The original (nonphysical) solution $\psi$ for the superposition constants
$a_1=a_2=1$ and the strictly isospectral solutions for the same values of
$\lambda$ as in Fig. 1.

{\bf Fig. 3}: 
 The flux solutions $\phi$ corresponding to the $\psi$ ones in Fig. 2.
\underline{Unpublished note}: 
In an inverted potential the isospectral `flux' solutions 
could be interpreted as a sort of `resonant' states.

\newpage
\centerline{
\epsfxsize=260pt
\epsfbox{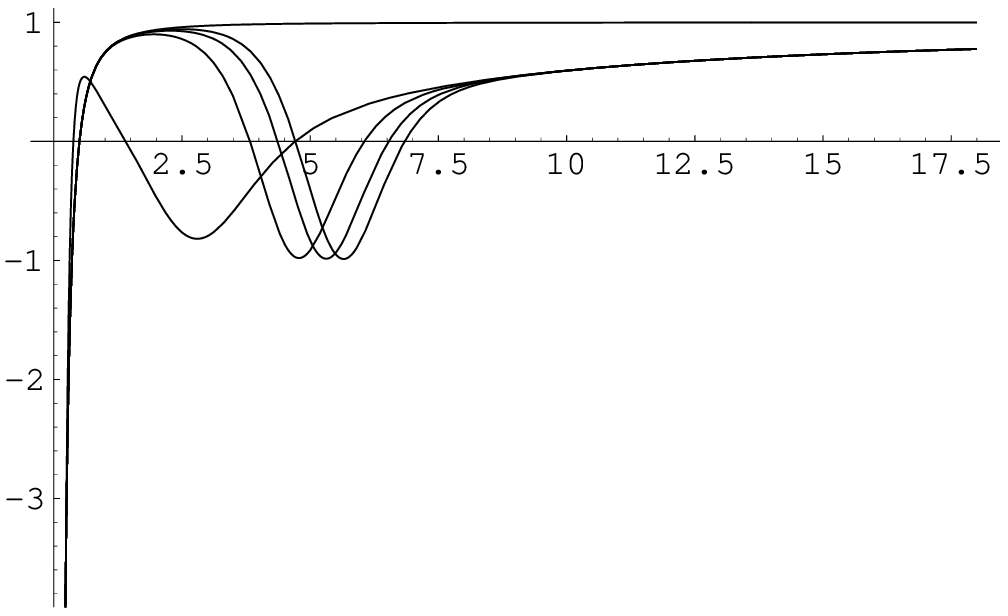}}
\vskip 4ex
\begin{center}
{\small{Fig. 1}\\
}
\end{center}

\vskip 2ex
\centerline{
\epsfxsize=280pt
\epsfbox{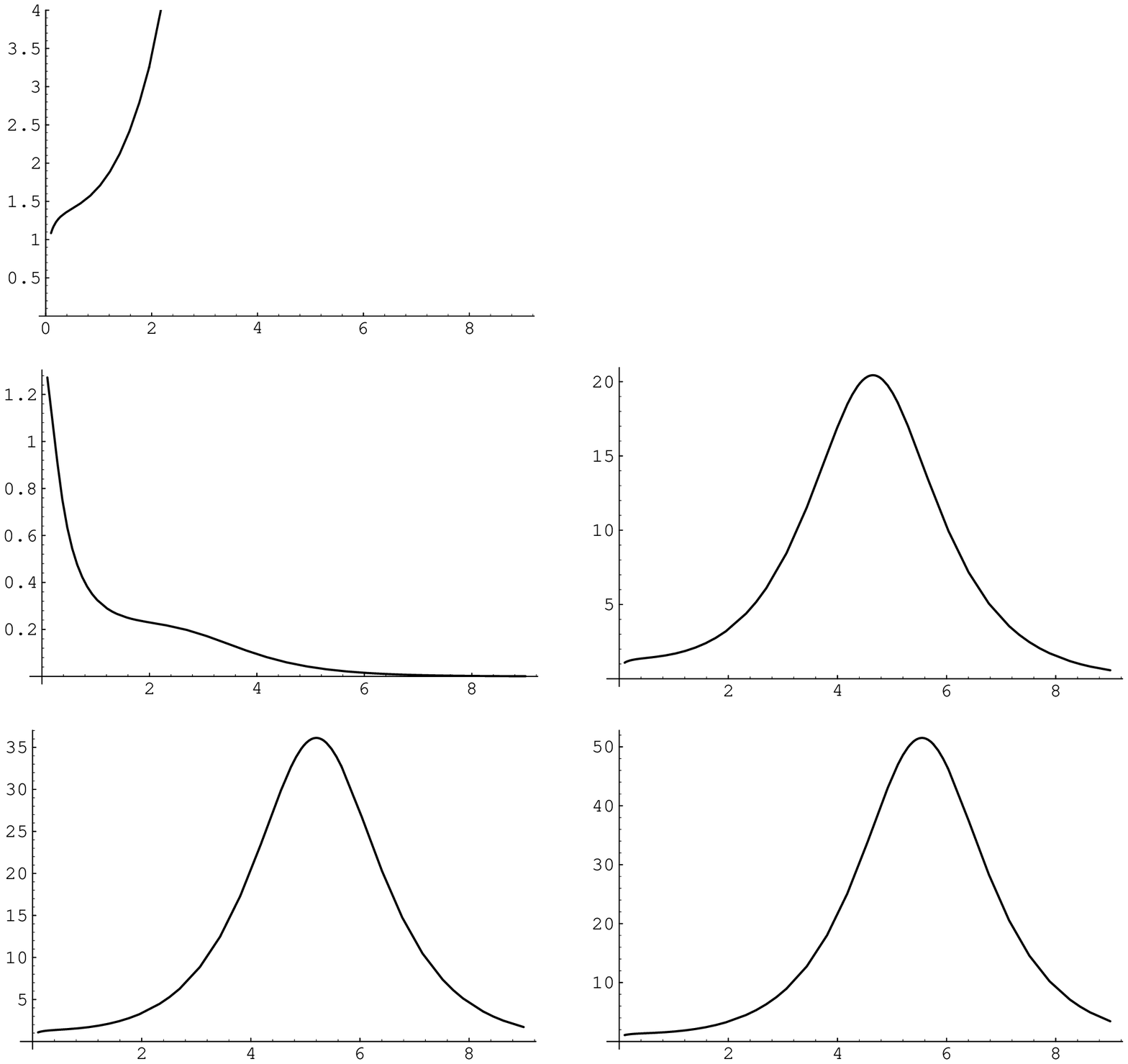}}
\vskip 4ex
\begin{center}
{\small{Fig. 2}\\
}
\end{center}

\vskip 2ex
\centerline{
\epsfxsize=280pt
\epsfbox{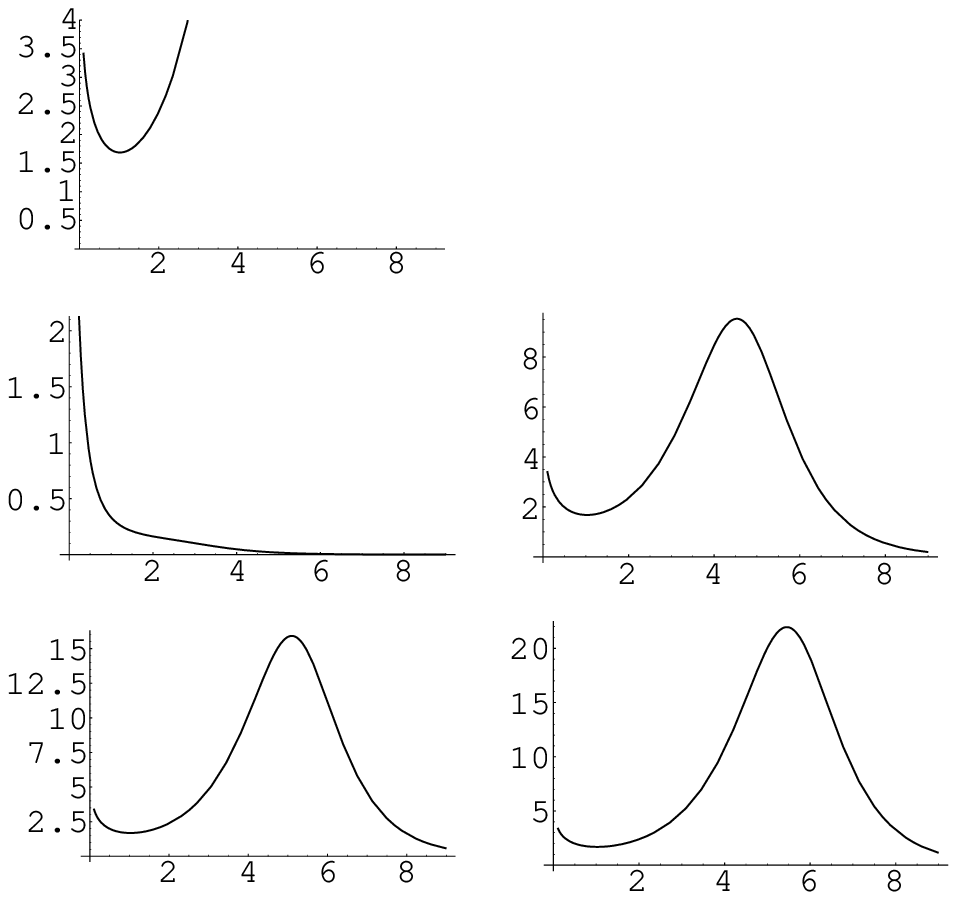}}
\vskip 4ex
\begin{center}
{\small{Fig. 3}\\
}
\end{center}

\end{document}